# Comparative study of the evolution of human cancer gene duplications across fish

Ciara Baines[1]*, Richard Meitern[1]*, Randel Kreitsberg[1], Tuul Sepp[1]

[1]Institute of Ecology and Earth Sciences, University of Tartu, Vanemuise 46, 51014 Tartu, Estonia

*Corresponding authors: ciara.baines@ut.ee, richard.meitern@ut.ee

## Abstract

Comparative studies of cancer-related genes allow us to gain novel information about the evolution and function of these genes, but also to understand cancer as a driving force in biological systems and species life histories. So far, comparative studies of cancer genes have focused on mammals. Here, we provide the first comparative study of cancer-related gene copy number variation in fish. As fish are evolutionarily older and genetically more diverse than mammals, their tumour suppression mechanisms should not only include most of the mammalian mechanisms, but also reveal novel (but potentially phylogenetically older) previously undetected mechanisms. We have matched the sequenced genomes of 65 fish species from the Ensemble database with the cancer gene information from the COSMIC database. By calculating the number of gene copies across species using the Ensembl CAFE data (providing species trees for gene copy number counts), we were able to develop a novel, less resource demanding method for ortholog identification. Our analysis demonstrates a masked relationship with cancer-related gene copy number variation (CNV) and maximum lifespan in fish species, suggesting that higher tumour suppressor gene CNV lengthens and oncogene CNV shortens lifespan, when both traits are added to the model. Based on the correlation between tumour suppressor and oncogene CNV, we were able to show which species have more tumour suppressors in relation to oncogenes. It could therefore be suggested that these species have stronger genetic defences against oncogenic processes. Fish studies could yet be a largely unexplored treasure trove for understanding the evolution and ecology of cancer, by providing novel insights into the study of cancer and tumour suppression, in addition to the study of fish evolution, life-history trade-offs, and ecology.

## Introduction

Cancer is a disease that arose with multicellularity and is caused by a variety of mutations that occur either somatically, arising throughout the organism's lifetime, or are inherited through the germline (Trigos et al. 2018). It is estimated that approximately 90% of mutations leading to cancer in humans are caused from somatic mutations (Sondka et al. 2018). Evolution, as a result, has led to the selection of genes that reduce the risk of an organism to neoplastic development. It is understood that oncogenes (OGs), tumour-suppressor genes (TSGs), and differentiation genes are amongst the oldest gene classes in humans (Makashov et al. 2019), opening a possibility for gaining novel information about the evolution and function of these genes from comparative studies. Moreover, comparative studies allow us to understand that cancer is not only a disease but also a driving force in biological systems and species life histories (Nunney et al. 2015). Theoretically, species with longer lifespans or larger body size should be at a greater risk of cellular mutations that increase cancer risk due to a greater number of cellular divisions. However, genetic controls on neoplastic cellular proliferation vary between species, resulting in a lack of correlation between body size and cancer prevalence, a paradigm known as Peto's paradox (Peto et al. 1975; Caulin and Maley 2011; Tollis et al. 2017). These genetic controls include the upregulation or duplication of TSGs and the downregulation of OGs within an organism. TSGs can control potentially carcinogenic mutations through various mechanisms including apoptosis, cell cycle arrest, and senescence (Kumari et al. 2014). They can be divided into two major categories, known as caretakers and gatekeepers; caretaker genes control the maintenance of the genetic information integrity in each cell, whilst gatekeepers are genes that directly regulate tumour growth, codifying for proteins which either stimulate or inhibit proliferation, differentiation or apoptosis (Weitzman 2001).

Gene duplication is considered an important mechanism for creating genetic novelty, as it has contributed to the evolution of developmental programmes, the plasticity of a genome, and the capability of a species to adapt to changing environments (Magadum et al. 2013). It has been suggested that increased copy numbers of TSGs are amongst the most effective routes to enhanced cancer resistance (Vazquez and Lynch 2021). Furthermore, duplicated TSGs can sometimes be selectively lost, which could be a macroevolutionary route towards lower cancer resistance (Glenfield and Innan 2021). TSG duplication is one of the possible mechanisms behind increased cancer resistance in large-bodied and/or long-lived mammals. For example, low cancer mortality rates in elephants (Proboscidean lineage) may be linked to 20 genomic copies of the gene TP53 (Abegglen et al. 2015; Sulak et al. 2016), a tumour suppressor responsible for apoptosis, senescence, and cell cycle arrest in the presence of damaged DNA (Kumari et al. 2014). In blind mole rats (*Spalax sp.*), another tumour suppression mechanism has evolved, through duplications of genes in the interferon pathway, leading to interferon- mediated concerted cell death, a strategy that has been proposed to counteract the weakened pro- apoptotic function of the p53 protein (Gorbunova et al. 2012). A recent study in cetaceans indicated positive selection within the CXCR2 gene, an important regulator of DNA damage, tumour dissemination and immune system, and 71 duplicated genes, had roles such as the regulation of senescence, cell proliferation and metabolism (Tejada-Martinez et al. 2021). Another recent study focusing on the evolution of elephants and their relatives (Proboscideans) from their smaller-bodied ancestors (Afrotherians) indicated that tumour suppressor duplication was pervasive in Afrotherian genomes, suggesting that duplication of TSGs facilitated the evolution of increased body size (Tejada-Martinez et al. 2021).

Another side of the TSG coin are OGs, genes that encode proteins that can induce cancer in animals (Lodish et al. 2000), and produce transcription factors, chromatin remodelers, growth factors, growth factor receptors, signal transducers, and apoptosis regulators (Croce 2009). Of the many known OGs, all but a few are derived from normal cellular genes called proto-oncogenes, whose products participate in cellular growth-controlling pathways (Lodish et al. 2000), by encoding proteins that stimulate cell division, inhibit cell differentiation, and halt cell death (Chial 2008). All these processes are important for normal development and maintenance of tissues and organs. Due to their basic role in animal life, proto-oncogenes have been highly conserved over eons of evolutionary time (Lodish et al. 2000). For growing big and/or living long, an increased number/function of proto-oncogenes is expected, bringing along the risk of these genes turning into OGs by a gain-of-function mutation. This risk can be counteracted by an increase in the number of (copies of) TSGs. Whilst comparative studies have so far mainly focused on TSGs, a strong correlation between the number of TSGs and (proto-)oncogenes is expected and has been demonstrated on the between-species level in mammals (Tollis et al. 2020). We suggest that instead of focusing on the TSGs in comparative studies, a balance between TSGs and OGs should be considered. For example, it is possible that a species with a lower number of TSGs is still more resistant to cancer due to a lower number of OGs. Having fewer proto-oncogenes decreases the chances of an oncogenic mutation and therefore decreases the overall probability of cancer development (Caulin and Maley, 2011). However, since proto-oncogenes are serving important functions in cellular growth, development, and cellular maintenance, eliminating them is costly in terms of growth and longevity. Accordingly, species with larger bodies and longer lifespans would be expected to not just have more TSGs, but to have more TSGs relative to their number of (proto-)oncogenes, to counteract increased cancer risk.

It has now been widely accepted that using a wider variety of model species provides novel insights into the genetic basis of disease resistance, and allows a better understanding of tumour suppression mechanisms, evolutionary and ecological importance of oncogenic processes, and of the link between modern environment and cancer (Pesavento et al. 2018; Giraudeau et al. 2018, Hamede et al. 2020). Wildlife cancer studies are therefore an emerging research topic, with the potential for deepening our understanding of the mechanisms that reduce cancer risk in some species, potentially leading to novel ideas for cancer treatments. Depending on their longevity, body size, life history strategy, but also environmental (oncogenic) pressures, species should deploy different tumour suppression strategies. However, to date, comparative studies of tumour suppression mechanisms have focused on mammals (e.g. Abegglen et al. 2015; Tollis et al. 2017; Seluanov et al. 2018; Tejada-Martinez et al. 2021; Vazquez and Lynch 2021; Yu et al. 2021). This focus should be widened to include other vertebrate groups, to verify the patterns seen in mammals in another phylogenetic and ecological groups, and to discover novel mechanisms and evolutionary directions related to the evolution of tumour suppression. Fish are evolutionarily older and genetically more diverse than mammals (Buchmann 2014) and accordingly, their tumour suppression mechanisms should not only include most of the mammalian mechanisms, but also reveal novel (but potentially phylogenetically older) previously undetected mechanisms. The life-history trade-offs that have shaped the relationships between tumour suppression and traits such as longevity, body size, and fecundity, should be more apparent in an evolutionarily more diverse group. In addition, there is evidence that fish lineages have evolved increased rates of duplicated genes than mammals (Robinson-Rechavi and Laudet 2001), suggesting a possibility that tumour suppression and gene duplications could be related to life-history more reliably in fish compared to mammals. Given these differences in evolutionary history between vertebrate groups, expanding our knowledge of how cancer defences have evolved across a wider range of taxa could deepen our understanding of cancer as a disease, and the evolutionary processes that increase and decrease a species risk of neoplastic development.

We expand on the comparative study undertaken by Tollis et al. (2020) on mammals to better understand the relationship between life history traits and phylogenetic data on known TSGs and OGs. Tollis et al. (2020) demonstrates a strong correlation between the TSG and OG normalized copy numbers that possibly indicates some sort of evolutionary constraint holding a genomic balance between TSGs and OGs. Both types of genes perform important tasks in retaining homeostasis. Arguably, the most notable role of these genes is to regulate growth. Indeed, any of the genes responsible for increased body size are also OGs, as larger individuals generally have higher rates of cancer within the species (Nunney 2018). Likewise, TSGs have the role of reducing cell proliferation. Nevertheless, not all genes contributing to body size are TSGs or OGs. Furthermore, body size (larger individuals benefit from reduced predation rates) and cancer susceptibility are just two among many factors affecting animal lifespan. Based on these premises we hypothesize that the lifespan of fish species is correlated positively to the copy numbers of TSGs and negatively to copy numbers of OGs when correcting for body size (figure 1). To test this hypothesis, we have conducted a comparative analysis that examines the relationship between life history traits (longevity, body size) and the number of cancer-related gene duplications in fish. Using the Catalogue of Somatic Mutations in Cancer (COSMIC; Sondka et al. 2018), we estimated the copy numbers of human cancer gene homologs in 85 complete genomes from across the phylogenetic tree of aquatic vertebrates (except mammals). The COSMIC database contains cancer related genetic data specifically based on the human genome. As many human genes have orthologues in other vertebrate species, the human cancer genes provide a reasonable proxy for testing hypotheses about life many human genes have orthologues in other vertebrate species (e.g. zebrafish (*Danio rerio*) have 70% overlap, Howe et al. 2013) and we are not aware of any collection of fish cancer genes, the human cancer genes provide a reasonable proxy for testing our hypothesis.

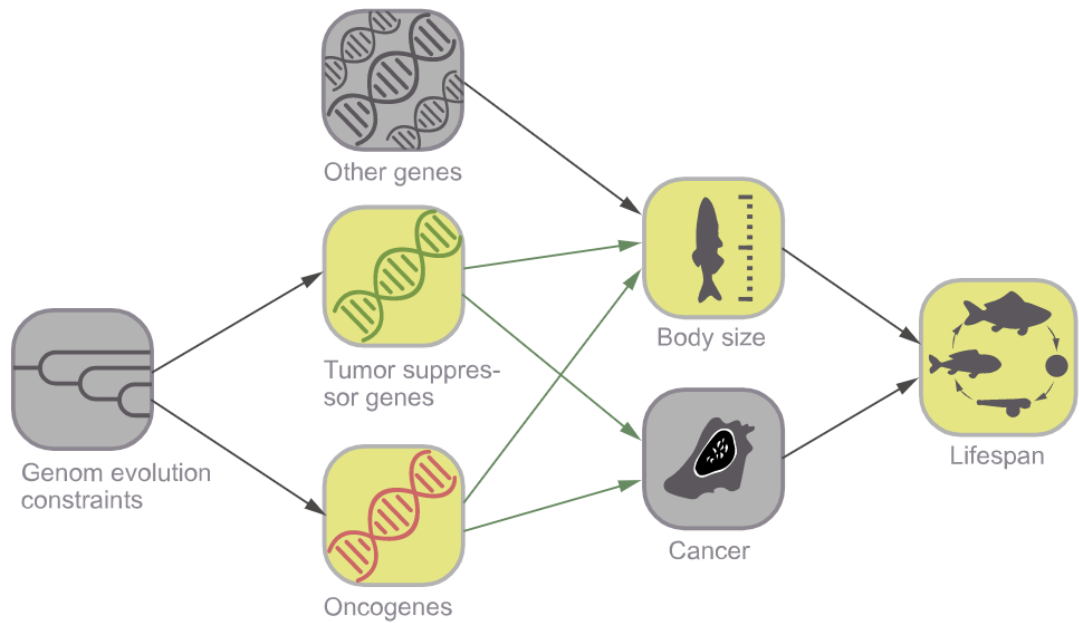

**Figure 1.** A directed acyclic graph depicting a simplified view of how lifespan may be influenced by copy numbers of TSGs and OGs. Gray boxes represent unobserved and green boxes represent observed variables. Arrows indicate possible causal pathways. Green arrows indicate causal paths to be tested.

## New approaches

We are using the Ensembl orthology database and the Ensembl CAFE in a novel context. We calculate the number of gene copy numbers across species using these databases instead of performing pairwise or multiple genome alignment for ortholog identification. This allows for a less resource demanding analysis pipeline for identifying copy numbers of genes across species.

## Methods

The list of cancer related genes used is obtained from COSMIC (Sondka et al. 2018). Two sub-lists of the COSMIC database are used for copy number variation (CNV) analyses. The first is the COSMIC tier 1 genes (comprising only of genes with strongly established link between mutations and cancer) and the second included both COSMIC tier 1 and tier 2 genes. Tier 2 genes have indications of a role in cancer with less extensive evidence than tier 1. COSMIC is a manually curated list of human cancer genes, that also assigns genes as either TSGs or OGs. COSMIC also provides the mutation type such as germline, somatic, or both. In addition, we classified each TSG as being a gatekeeper gene or a caretaker gene according to the list provided by Tollis et al. (2020). For detailed code see supplementary material S1.

The copy number count of cancer related genes, in fish, in this paper is obtained differently from Tollis et al. (2020). In order to get the copy number of the above-mentioned COSMIC genes in different fish species two approaches were used. The first method included downloading the Ensembl CAFE (Herrero et al. 2016) species trees for all the COSMIC genes. The Ensembl CAFE provides species trees for gene copy number counts. In Ensembl CAFÉ, gene gain and loss data is estimated from the number of gene copies whilst also taking into account the lineage information (De Bie et al. 2006, Herrero et al. 2016). The second method included downloading the list of human COSMIC gene orthologs for each species represented in the Ensembl database using BioMart (Kinsella et al. 2011) and counting the unique confident orthologs in each species for each gene. Both chosen approaches are computationally much less intensive than that used by Tollis et al. (2020) but provide similarly reliable CNV data as the BLAST/BLAT approach (Kent 2002, used in Tollis et al. 2020). This enables re-running the analysis whenever the Ensembl databases are updated. The exact details of how the CAFE and ortholog gene counting methods are implemented are accessible from the supplementary material S2 and the functions therein.

The normalized copy number counts for both cancer gene lists (COSMIC Tier 1 and COSMIC Tier 1&2) and for both copy number count methods (CAFE and Ortholog) was calculated according to Tollis et al. (2020). Both the TSG and OG counts were implemented so that genes that have been classified as both TSGs and OGs in the COSMIC database were excluded from calculation of CNV (see S2). For validation of calculation steps on the Tollis et al. (2020) dataset see supplementary materials S3.

The maximum length and lifespan data (as well as other parameters) where obtained mainly from FishBase (Froese and Pauly 2021) and AnAge database (Tacutu et al. 2017). For some species where lifespan and body size data was not available from either Fishbase or AnAge databases, we looked for reliable data in articles and other sources. Species with no maximum lifespan data were excluded from the dataset. The longevity quotient (LQ) was calculated according to Tollis et al. (2020).

The phylogenetic tree for the fish species together with branch lengths was obtained from timetree.org (Kumar et al. 2017). Species that were missing in the timetree.org database were excluded from the analysis as phylogenetically informed regressions cannot be done without phylogenetic distances. The body size (maximum body length) and lifespan was log transformed prior to analysis. The normalized CNV counts were standardized (i.e. converted to z-scores) prior to all analyses. All the statistical analysis was performed in R (version 4.0.5, R Core Team 2021) using the caper package (Orme et al. 2013) for phylogenetically informed regressions. If $\lambda$, $\kappa$ and $\delta$ values are provided the branch lengths were optimized using maximum likelihood. For more details on $\lambda$, $\kappa$ and $\delta$ see the caper package manual (Orme et al. 2013). Otherwise $\lambda$, $\kappa$ and $\delta$ values were fixed at 1. Other used packages included base, utils, stats, (R Core Team 2021) ggplot2 (Wickham et al. 2016), ggtree (Yu 2020), tidytree (Yu 2021), biomaRt (Durinck et al. 2009), ape (Paradis and Schliep 2019), AnnotationDbi (Pagès et al. 2019), dagitty (Textor et al. 2016) and dependencies within those.

## Results

**Lifespan vs longevity**

For 12 of the 65 fish species (*Actinopterygii*), we were unable to recover maximum lifespan data. For 25 we found this data from AnAge, for 12 from Fishbase and 10 from articles. The remaining data (9 species) is distributed between 5 different less reliable sources (see S6). For the 4 fishlike aquatic species (see S2 for clarification), 3 had age and size data available in Fishbase or AnAge. Nevertheless, from hereon we will present results on the full dataset whilst the results using only fish species from class *Actinopterygii* are presented in S5. When branch lengths were optimized using maximum likelihood, maximum lifespan was related to maximum body size (Figure 2). At fixed branch lengths the relationship holds only for data from reliable sources (see S5). This is true for average age as well (S5).

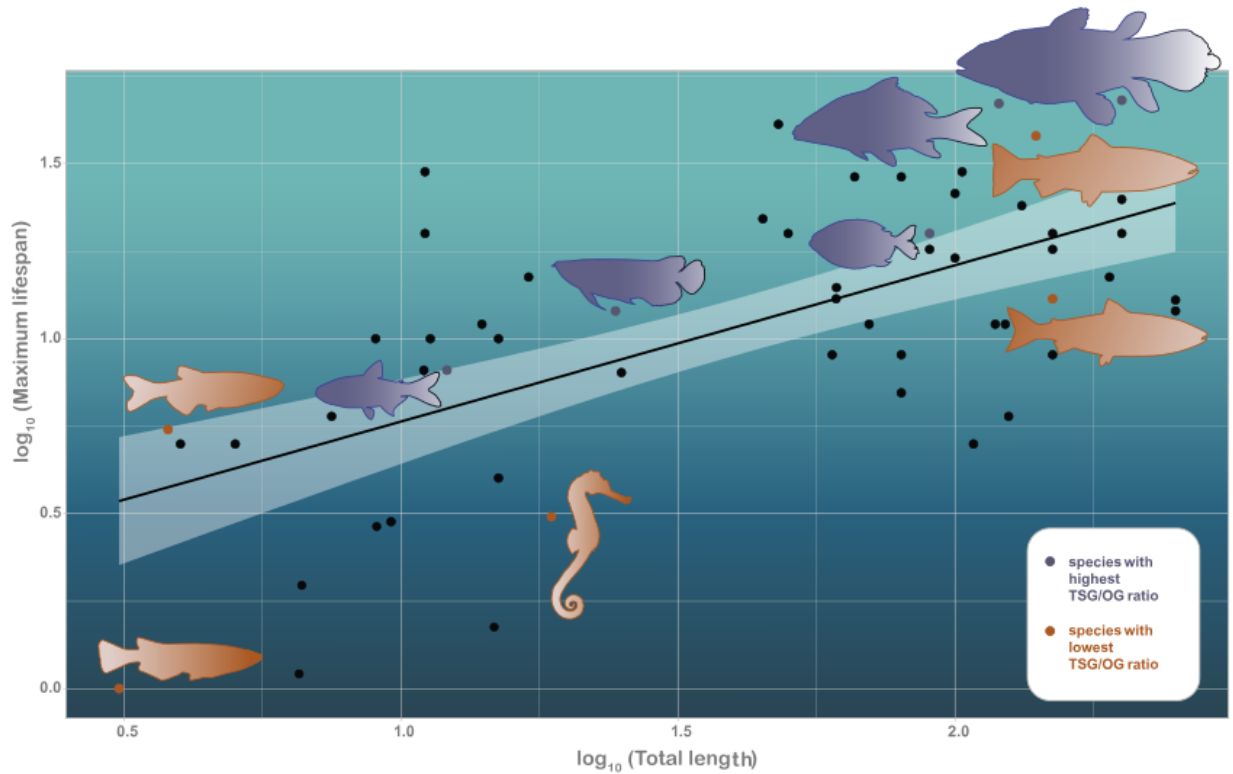

**Figure 2.** Linear regression between log transformed maximum body length and maximum lifespan. Each point in the plot represents a species. The line and the confidence intervals depicted in the plot come from standard linear regression, the values $R^2$, p and N are from phylogenetically adjusted regression, where λ, κ and δ are optimized using maximum likelihood. Purple dots and associated silhouettes illustrate 5 species with highest tumor suppressor gene CNV to oncogene CNV ratio (*Latimeria chalumnae, Cyprinus carpio, Pygocentrus nattereri, Scleropages formosus, and Astyanax mexicanus*), orange dots 5 species with lowest ratio (*Salmo salar, Danio rerio, Hippocampus comes, Salmo trutta, Oryzias sinensis*) (only including species for whom the lifespan is known).

**Human cancer gene duplications in fish genomes**

We queried 68 genome assemblies representing three clades, Actinopterygii (ray-finned fish, 65 species), Cyclostomata (jawless fish 3 species), and Sarcopterygii (fringe-finned fish, 1 species)

(fig. 3) for 715 human cancer genes. Altogether the COSMIC list holds 243 pure TSGs, 243 pure OGs, 72 genes classified as both, 134 classified as pure fusion genes (i.e. genes resulting in cancer if translocated) and 31 genes as all (OGs, TSGs and fusions). We obtained normalized copy number counts for two cancer gene lists (COSMIC Tier 1 and COSMIC Tier 1&2, Tate et al. 2019), using copy number count methods that take into account the lineage information (CAFE, Herrero et al. 2016). As all species diverged from the lineage leading to humans at the same time point, we did not need to test for the potential systematic bias in our ability to identify human cancer genes in nonhuman genomes, as was done in the analogous comparative analysis of mammalian genomes (Tollis et al. 2020). From all queried human cancer genes, an average of 218 (±11 SD) TSG and 192 (±12 SD) OG orthologs were identified in these species using the CAFE approach and 170 (±31 SD) TSG and 152 (±27 SD) OG orthologs for the ortholog approach (see S2 for numbers for all subsets). The methodology of obtaining copy number counts in this paper provides similar results to the methodology of Tollis et al. (2020, see S3). In addition, the different COSMIC gene CNVs (TSGs, OGs etc.) correlate positively ($R > 0.3$) regardless of the method used to obtain copy number counts (CAFE vs ortholog) or subsets of cancer genes (COSMIC tier 1 vs COSMIC tier 1&2, see S5). However, the total number of species in the analysis is larger for CAFE as the ortholog approach failed to produce copy number counts for some species (S5).

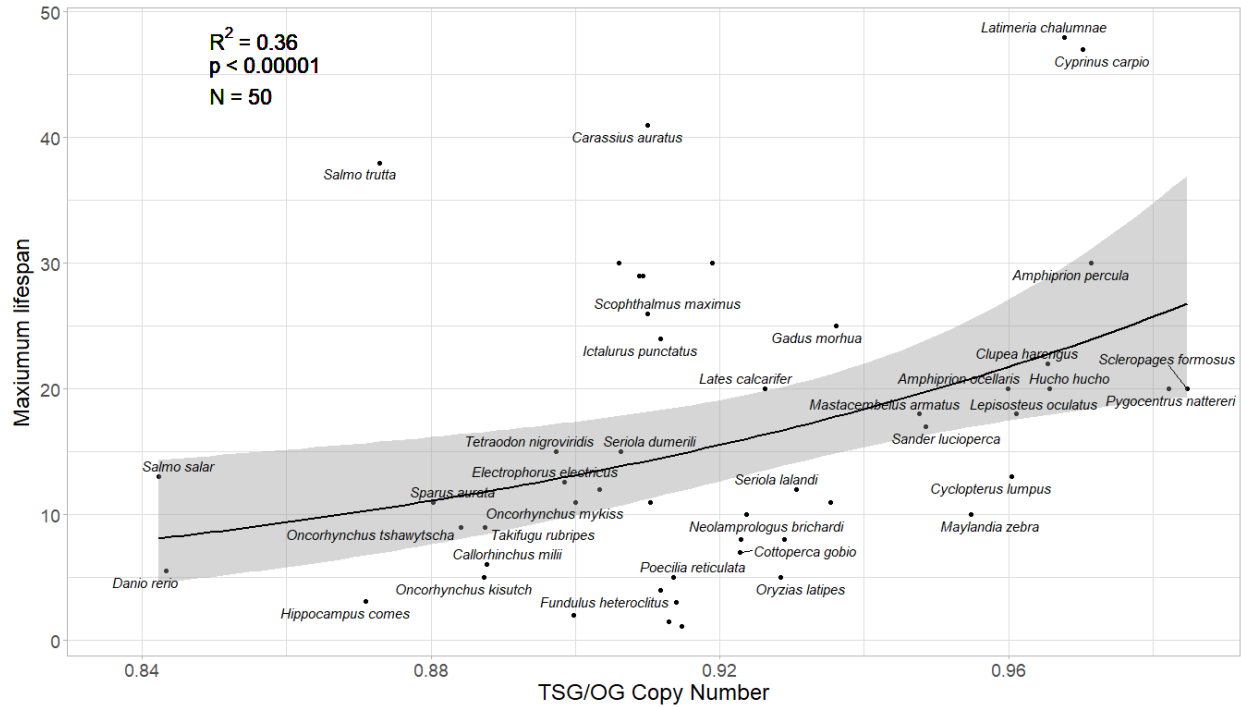

**Figure 3.** Linear regression between maximum lifespan and the normalized count of TSG, divided by the normalized count of OGs obtained from the CAFE approach and including only COSMIC Tier 1 genes. Each point in the plot represents a species in the dataset. The line and the confidence intervals depicted in the plot come from a log linked general linear model (i.e. not adjusted phylogenetically), the values $R^2$, p and N are from phylogenetically adjusted linear regression where the maximum lifespan is log transformed. The λ, κ and δ values are fixed at 1.

**Tumour suppressor genes balance oncogenes**

We found a strong correlation between the number of TSGs (all TSGs, gatekeeper genes and caretaker genes) and the number of OGs in studied genomes (fig. 4). Phylogenetically adjusted regressions for all TSGs were $R^2=0.93$, $p < 0.00001$, for gatekeepers $R^2=0.93$, $p < 0.0001$, and for caretakers $R^2=0.43$, $p < 0.00001$ (N=59 for all comparisons). All these results (and most of the results described in the next section) remained significant when we removed the two fish families with an extra round of whole genome duplications (*Salmonidae* and *Cyprinidae*, please see discussion for more information) from the analysis (see Supplementary materials 7 file for analysis results without these two families).

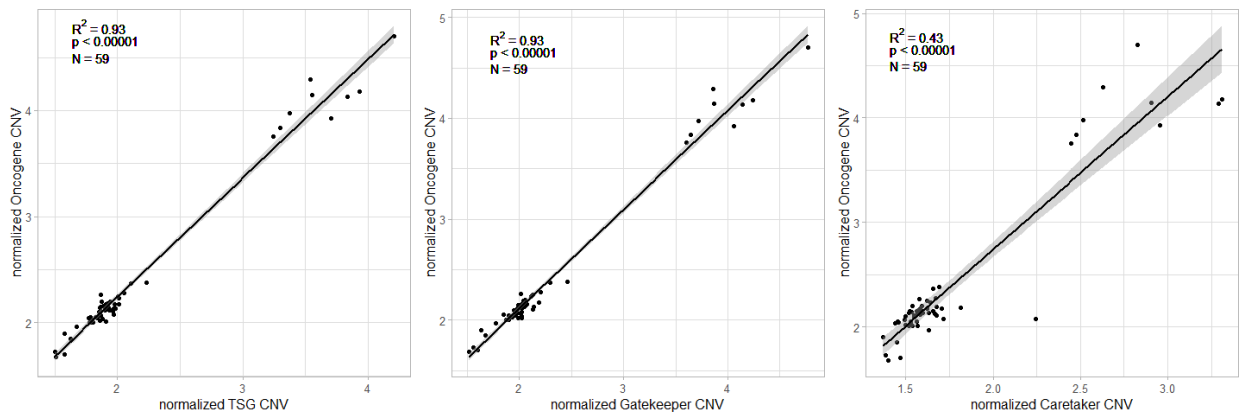

**Figure 4.** Linear regression between normalized copy number variations (CNV) of different subsets of TSGs (all TSGs, gatekeeper genes and caretaker genes) and OG CNV. The CNVs have been obtained using the CAFE approach and both COSMIC Tier 1 & 2 genes are included. Each point in the plot represents a species. The line and the confidence intervals depicted in the plot come from standard linear regression, the values $R^2$, p and N are from phylogenetically adjusted regression. The λ, κ and δ values are fixed at 1.

**Tumour suppressor genes lengthen, oncogenes shorten lifespan**

The magnitude of lifespan is positively related to the total number of TSGs and negatively to total number of OGs irrespective of branch length optimization (Table 1, Figure 3), the inclusion or exclusion of body size, or low quality data points in the model (see S5). However, the relationship reveals itself only when both OG and TSG counts are included in the model. The same result, a masking relationship between TSG and OGs, also holds true for another measure of lifespan: the longevity quotient (LQ) (see S5). To test if the same masked relationship is present in the mammalian dataset, we ran a comparable analysis with mammals. In the mammalian dataset, we could not reveal the masked relationship between lifespan and TSGs or OGs (see S4).

**Table 1.** Results from the phylogenetically adjusted regression with log maximum lifespan as the response variable. The predictors are body size (log maximum total length) and normalized TSG and OG copy number counts (CNV). The $\lambda$, $\kappa$ and $\delta$ values are either fixed at 1 (left) or maximum likelihood optimized (right). The CNVs have been obtained using the CAFE approach and only COSMIC Tier 1 genes are included. All species are included (N=50).

|  | Fixed branch lengths | | | | optimized using maximum likelihood | | | |
| --- | --- | --- | --- | --- | --- | --- | --- | --- |
|  | Estimate | SE | t | p | Estimate | SE | t | p |
| (Intercept) | 0.83 | 0.65 | 1.27 | 0.21 | 0.44 | 0.14 | 3.11 | 0.0033 |
| Body size | 0.19 | 0.16 | 1.23 | 0.22 | 0.36 | 0.08 | 4.72 | <0.0001 |
| OG CNV | -1.18 | 0.34 | -3.5 | 0.001 | -0.85 | 0.27 | -3.08 | 0.0035 |
| TSG CNV | 1.30 | 0.34 | 3.88 | 0.0003 | 0.89 | 0.28 | 3.21 | 0.0024 |
|  | $\kappa = 1$ | $\lambda = 1$ | $\delta = 1$ | | $\kappa = 0.15$ | $\lambda = 0$ | $\delta = 0.52$ | |

**Species-specific differences**

We found that many human cancer genes are indeed also duplicated in fish genomes. CNV varied between species, as did the ratio of TSGs/OGs (fig. 5). As expected, the species from the fish families that had undergone an extra round of whole genome duplication (*Salmonidae* and *Cyprinidae*) stand out as species with the highest copy numbers of TSGs and OGs. However, even within the fish species with smaller genomes, TSG and OG CNV ranges from 1.5 to 2.2. When looking separately at fish species outside the salmonid and cyprinid families, species with highest copy numbers of TSGs are two tropical fish, Asian arowana (*Scleropages formosus*) and mormyrid electric fish (*Paramormyrops kingsleyae*), and one temperate fish, the ballan wrasse (*Labrus bergylta*) (based on COSMIC tier 1 gene list, which is more reliable in regards of links of genes with cancer compared to the full list). As TSG and OG copy numbers are correlated, we also calculated the TSG/OG ratio for all studied species (fig. 5), with the suggestion that species with the highest ratio invest more into cancer defences compared to species with the lowest ratio. Since this approach compensates for the whole genome duplication in two fish families, we can make comparisons across all studied species. According to this calculation, three species with the highest TSG/OG CNV ratio were blind cave tetra (*Astyanax mexicanus*, TSG/OG copy number ratio 1.017), Asian arowana (0.985), and the red-bellied piranha (*Pygocentrus nattereri,* 0.982). Three species with the lowest TSG/OG copy number ratio were zebrafish (0.843), Atlantic salmon (*Salmo salar*, 0.842), and reedfish (known also as ropefish, *Erpetoichthys calabaricus,* 0.837).

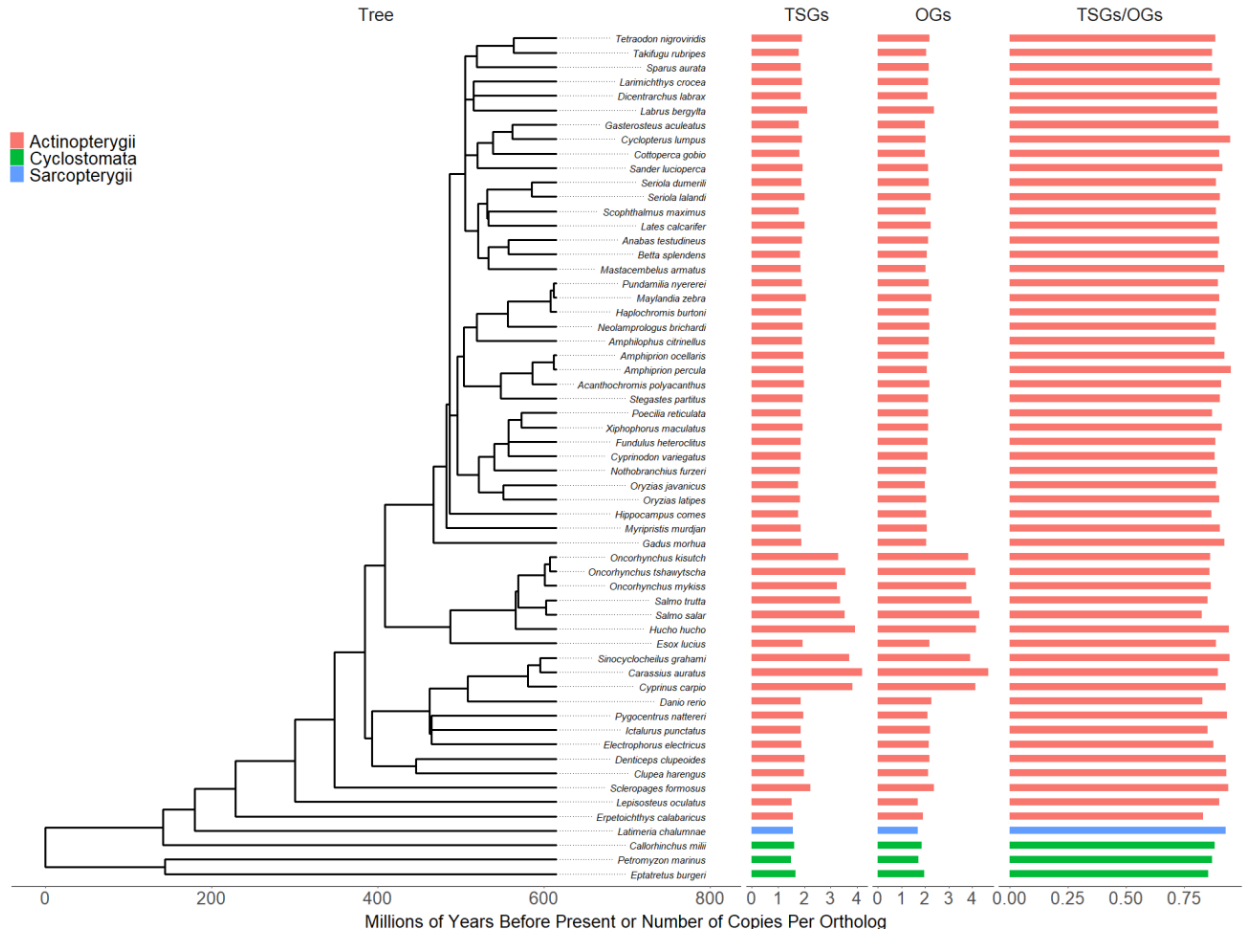

**Figure 5**. The species tree with branch lengths and counts for each species for normalized TSGs and OGs. TSGs/OGs is the ratio between normalized TSG counts and normalized OG counts.

## Discussion

To date, comparative studies that have analysed cancer-related gene duplications have been done on mammalian genomes (Tollis et al. 2020; Tejada-Martinez et al. 2021; Vazquez and Lynch 2021) and have suggested a link between lifespan and tumour suppressor gene copy numbers. Focusing on a phylogenetically older and genetically more diverse class of vertebrates could provide a control for the generalizability of the detected patterns but can also reveal patterns and trade-offs that are not present in mammalian genomes. Here, we have provided the first comprehensive survey of cancer-related gene duplications across the fish radiation, incorporating 715 human cancer genes with known orthologues in the genomes of 68 species.

Whilst reading the following discussion of the results, it is important to keep in mind that the dataset behind the analysis is far from perfect. We are using the best available database of known cancer genes, but since studies on wildlife cancer are still in their infancy, using a human database is the only available option for understanding the link between cancer related genes and species' life history. Since proto-oncogenes and TSGs are generally phylogenetically old, some of them dating back to the emergence of multicellularity (Lodish et al. 2000, Makashov et al. 2019), it is reasonable to make this link with the human database in the absence of a more taxonomically wide database. However, we hope that future studies of wildlife cancer and genomics will soon result in a novel "wildlife COSMIC" database in which the analyses described here could be verified. Until it has been experimentally verified that human cancer genes in this study share the same function in other species, our results must be taken with caution. It is important to note that it is likely there are other fish-specific pathways for tumour suppression in addition to those causally linked to human cancers, which we will miss in our current analysis. However, compared to previously published studies in mammals using the same human-centred approach, our study benefits from the fact that the evolutionary distance from humans should not play a role in our comparative analysis on fish. As a novel approach, we used the Ensembl orthology database and the Ensembl CAFE to calculate the gene copy numbers across species. Unfortunately, although the Ensembl database is considered of good quality, it is still missing a substantial number of species that already have a sequenced genome available. We cannot exclude the possibility that adding other aquatic vertebrate species to our dataset would have a significant effect on the results.

Both TSGs and OGs predict lifespan in fish. Interestingly, the results suggest the existence of a masked relationship between TSGs and OGs. The correlations of both OG CNV and TSG CNV are not detectable when correlated with the maximum expected lifespan individually. However, when both are included in the model, both have a significant effect in predicting lifespan. We are fully aware that some fields teach that correlated predictors should not be included into a model. However, this should not be an issue for two reasons: firstly, pairwise associations are not a problem (McElreath, 2020) under our assumption that there are genomic constraints keeping the genome in balance and, secondly, the relationship holds for the TSG/OG ratio (Figure 3) and is insensitive to inclusion or exclusion of various subsets of species. As the directions are opposite (more OGs reduce and more TSGs increase lifespan), the effect is observable only when both are included within the model, further highlighting the existence of a masked relationship. A higher number of copies of TSGs has a positive effect on lifespan, whilst a higher number of copies of OGs has a negative effect on lifespan. This suggests that in order to achieve a longer lifespan, species must compensate for the number of cellular growth inducing proto-oncogenes through increasing the number of copies of TSGs. In a previous comparative analysis with mammalian species, both the copy numbers of both TSGs and OGs were found to be positively correlated with longevity, a result that the authors found somewhat paradoxical (Tollis et al. 2020). Our results suggest that a high number of (proto-)oncogene copies can indeed shorten lifespan and needs to be compensated for with a higher number of TSG copies. A correlation between TSG and OG copy number was found for both mammals (Tollis et al. 2020) and fish, supporting this conclusion. In the mammalian dataset, it is possible that the compensatory mechanism, causing a strong correlation between TSGs and OGs, hid the negative effect of OGs on lifespan. In our analysis using the same dataset as Tollis et al. (2020), we could also not reveal this masked relationship between lifespan and TSGs or OGs. As the supplementary material S3 indicates, the CAFE and ortholog computational methods provide somewhat different CNV estimates. It is possible that Ensembl CAFE approach of calculating the gene gains and losses is also superior to the approach of Tollis et al. (2020) as it takes into account the phylogenetic tree of animals in CNV calculation. It is interesting to note that if we only kept mammal species with a genome assembly available in Ensembl (having a genome in Ensembl may be considered as having a genome of rather good quality), we were able to indeed demonstrate the same masked relationship (TSGs lengthen and OGs shorten lifespan if both are in the model together) in mammals that we discovered in the fish dataset. One possible explanation why the masked relationship does not hold as strongly for mammals compared to fish is the relatively small phylogenetic distance between different mammal species, compared to the distance differences within fish species. It might be that such a relationship emerges only on a larger phylogenetic scale. Another possible explanation is that the cancer genes that have an ortholog in fish are the most conserved and/or more important in terms of lifespan. Accordingly, we could speculate that the masked relationship only reveals itself in the fish and not in the mammal dataset, as other less relevant and perhaps evolutionary more novel, cancer-related genes are included.

Similarly to mammals and birds (where a strong correlation exists between lifespan and body mass, Healy et al. 2014), fish that live longer generally have longer bodies. As in other vertebrate classes, some species of fish live longer than expected for their body size, and some live shorter lives compared to other species in similar size. TSGs and OGs might be part of the story behind this variation, keeping in mind that it is mostly ecological selection pressures that have shaped lifespans of species over evolutionary time (Healy et al. 2014). We did not find a strong relationship between body size and TSG or OG copy numbers in our study. Species that grow larger tend to have slightly more copies of TSGs and OGs (Fig 11 in S5), but this trend is weak, and in the case of OGs, non-significant. Indeed, there are other adaptive roles for proto-OGs in addition to growth, e.g., cellular maintenance and survival (Creek et al. 2018). Whilst the positive association between TSG copy numbers and body length is expected, it was also not found in the similar comparative analysis of mammals (Tollis et al. 2020). Although it is known that within species, cancer risk increases with body size (Nunney 2018), this relationship does not seem to hold on the between-species level, potentially because of the upregulation of cancer defense mechanisms (Caulin and Maley 2011). We do not yet have good cancer prevalence data for most fish species, so it is still early to make conclusions about the so-called Peto's paradox (no increase in cancer prevalence with increased body size, Peto et al. 1975) in fish, but our results suggest that it is certainly a promising future research direction. Indeed, as fish grow throughout their life, it is logical to assume that defence mechanisms against the cost of growth (e.g. potentially increased cancer risk) are even more pronounced in this vertebrate class compared to classes with finite growth.

Species that have most cancer gene duplications in our study are the species that have gone through more rounds of whole genome duplications. Whilst all teleost fish have gone through three rounds of whole-genome duplication (WGD), a fourth round of duplication has taken place in salmonids (the salmonid-specific autotetraploidization event), which occurred in the common ancestor of salmonids ~100 Mya (Lien et al. 2016). Whilst only one autotetraploidization event has occurred in the common ancestor of salmonids, polyploidization has evolved independently on multiple occasions in Cyprinids, a large fish family including for example species like common carp (*Cyprinus carpio*) and goldfish (*Carassius sp.*) (Xu et al. 2019). When we exclude these two families, the species with the highest number of TSG copies are two tropical species, Asian arowana and mormyrid electric fish, and one temperate fish, the ballan wrasse (based on

COSMIC tier 1 gene list, which is more reliable in regards of links of genes with cancer compared to the full list). All these species stand out, as they have been selected among the few fish species for which the genome has been sequenced. For example, the genome of the mormyrid was sequenced in order to understand the evolution and development of electric organs, and to identify candidate housekeeping genes related to electrogenesis (Gallant et al. 2017). We could speculate that these are the fish with strongest tumour suppression systems, similarly to elephants (*Loxodonta Africana*), naked mole rats (*Heterocephalus glaber)*, two-toed sloth (*Choloepus hoffmanni*) and nine-banded armadillos (*Dasypus novemcinctus*) in mammalian class (Tollis et al. 2020).

However, as we can see that TSG and OG copy numbers are correlated, we suggest that looking at TSG copy numbers in relation to OG copy numbers might be more informative in terms of cancer resistance and cancer susceptibility. This approach also allows the inclusion of salmonids and cyprinids in the discussion. Blind cave tetra is the species with the highest TSG/OG ratio in our dataset. This species has undergone a recent rapid evolutionary change, dividing into two subspecies, one that lives in total and permanent darkness and lacks eyes and pigmentation, the other an "ancestral" multi-coloured tropical freshwater fish. With cave colonization, this species has undergone strong selective pressure and extreme morphological evolution and can be used to understand the evolution of specific traits and genetic mechanisms that support rapid habitat-based evolutionary change (Torres-Paz et al. 2018). Whether stronger tumour suppression is one of these traits remains to be studied. Next in line is the Asian arowana, who is still among the top three, in absolute TSG copy numbers as well as TSG/OG copy number ratio. This endangered and highly valued ornamental species stands out among fish due to its late sexual maturation and unusually high level of parental care (Scott and Fuller 1976). High tumour resistance could therefore be considered as a trait related to slow life history (Boddy et al. 2020). The last of the three species with highest TSG/OG ratio is red-bellied piranha, another fish species for whom parental care has been described (Queiroz et al. 2010).

Three species with the lowest TSG/OG copy number ratio were zebrafish, Atlantic salmon, and reedfish. Zebrafish has become one of the most common model organisms in cancer research in recent decades, due to rapid development, ease of care, similarity of tumourigenesis to humans, and its well-studied genome (Stoletov and Klemke 2008). If the fast life-history of zebrafish is linked to higher cancer susceptibility, zebrafish might be a model organism that is more similar to mice than to humans in terms of the evolution of tumour suppression mechanisms. In addition to the Atlantic salmon, several other salmon species tend to have low TSG/OG ratio. We could speculate that this is also related to life history, as several salmon species are semelparous, breeding only once in their life. Reproduction in semelparous species can lead to rapid severe pathology known as reproductive death by various mechanisms, due to very high levels of reproductive effort and drastically lowered investment in self-maintenance (Gems et al. 2021). Reduced tumour suppression could be one part of this strategy of low self-maintenance investment and prioritisation of growth/reproduction. The species with the lowest TSG/OG ratio in our dataset was reedfish, a facultative air-breather with an elongated body and the ability to move in both aquatic and terrestrial environments (Sacca and Burggren 1982). It might be assumed that adaptation to two very different environments would also require strong tumour suppression mechanisms, but that does not appear to be the case for reedfish. Based on this finding, we could speculate that switching between terrestrial and aquatic environments, and various levels of oxygen, could be an environmental factor that suppresses oncogenic processes, rather than induces them. Indeed, in humans, it has been shown that a change in oxygen pressure (hyperbaric oxygenation) could inhibit tumour cell proliferation (Granowitz et al. 2005). Whether reedfish are indeed better protected against cancer due to changes in oxygen pressures, therefore being able to afford lower investment in genome-based tumour suppression mechanisms, remains to be studied. Whilst this field of research is still in its infancy - the number of fish species that have been sequenced is still small, and the link between gene copy numbers and cancer is based on human data – it already shows great promise in providing a better understanding of the evolution of tumour suppression mechanisms. From the life-history perspective, we can suggest that fish species with slow life-history might exhibit stronger genomic defences against oncogenic processes, whereas fish with semelparous mating systems could be less protected against cancer. This finding might also have applications in conservation, as it might be possible to predict which species could be more vulnerable to oncogenic environmental change (e.g. oncogenic pollution exposure (Baines et al. 2021)). In conclusion, we were able to demonstrate a masked relationship with CNV and maximum lifespan in fish species and can suggest that a higher TSG count is probably behind the increased lifespan in some species. This masked relationship only reveals itself in fish data, similar comparative analysis in mammals did not support this finding, which indicates that studying different wild animal groups could provide complementary information about the evolution of tumour suppression. As fish are evolutionarily older and more diverse group compared to mammals, it is intriguing to suggest that fish studies could be a yet largely unexplored treasure trove for understanding the evolution and ecology of cancer. This field of research is a two-way street: it could provide novel insights into the study of cancer and tumour suppression, but also to the study of fish evolution, life-histories, and ecology.